# Donor−Acceptor Pairs Near Silicon Carbide Surfaces

Anil Bilgin,* Ian N. Hammock, Alexander A. High, and Giulia Galli*

**ABSTRACT:** Donor−acceptor pairs (DAPs) in wide-bandgap semiconductors are promising platforms for the realization of quantum technologies, due to their optically controllable, long-range dipolar interactions. Specifically, Al−N DAPs in bulk silicon carbide (SiC) have been predicted to enable coherent coupling over distances exceeding 10 nm. However, their practical implementations require an understanding of the properties of these pairs near surfaces and interfaces. Here, using first-principles calculations, we investigate how the presence of surfaces influence the stability and optical properties of Al−N DAPs in SiC, and we show that they retain favorable optical properties comparable to their bulk counterparts, despite a slight increase in electron−phonon coupling. Furthermore, we introduce the concept of surface-defect pairs (SDPs), where an electron−hole pair is generated between a near-surface defect and an occupied surface state located in the bandgap of the material. We show that vanadium-based SDPs near OH-terminated 4H-SiC surfaces exhibit dipoles naturally aligned perpendicular to the surface, greatly enhancing dipole−dipole coupling between SDPs. Our results also reveal significant polarization-dependent modulation in the stimulated emission and photoionization cross sections of *V*-based SDPs, which are tunable by 2 orders of magnitude via the incident laser's polarization angle. The near-surface defects investigated here provide novel possibilities for the development of hybrid quantum-classical interfaces, as they can be used to mediate information transfer between quantum nodes and integrated photonic circuits.

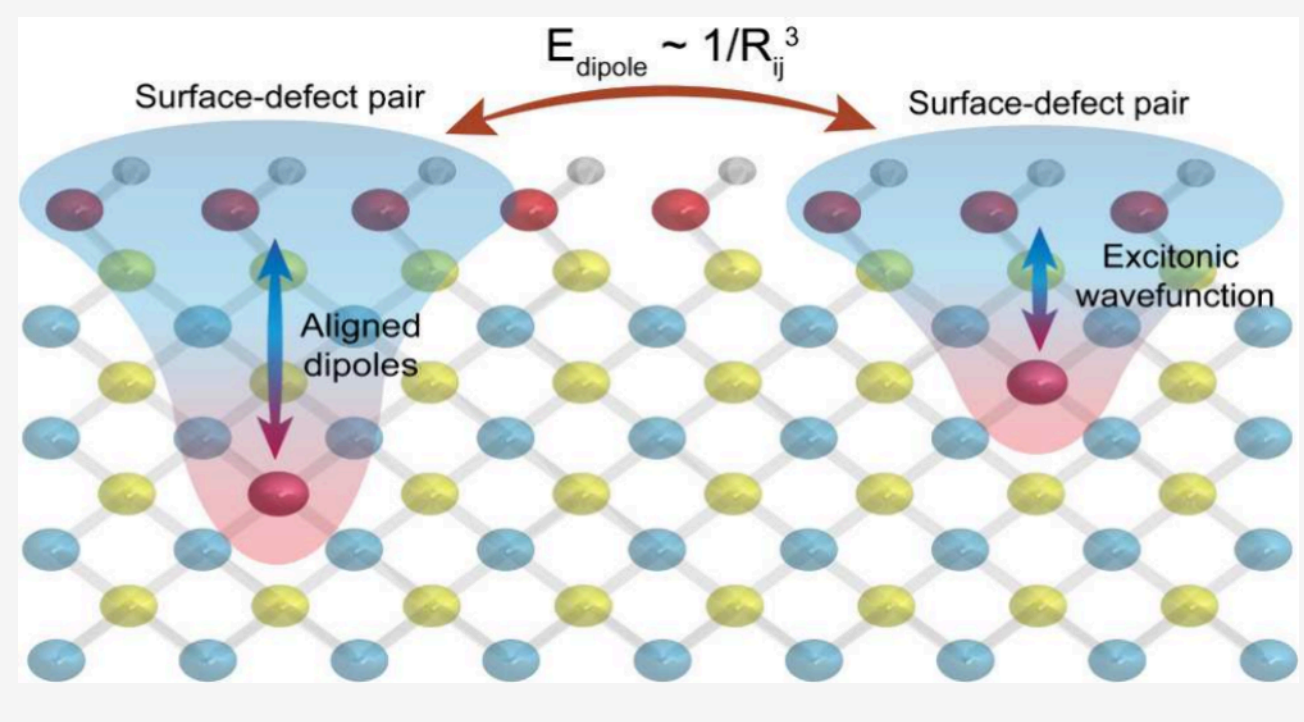
$E_{dipole} \sim 1/R_{ij}^3$
Surface-defect pair
Surface-defect pair
Aligned dipoles
Excitonic wavefunction

The development of solid-state quantum architectures relies on the ability to engineer and control quantum defects with long-range interactions and optical addressability.[1−4] Wide-bandgap semiconductors such as diamond[5−7] and silicon carbide (SiC)[8−10] have emerged as promising platforms for hosting defect-based qubits thanks to their favorable electronic and optical properties. In particular, donor−acceptor pairs (DAPs) across many platforms have been proposed as candidates for tunable quantum emitter arrays[11,12] and for their optically controllable, long-range dipole−dipole interactions.[13] The latter offers a mechanism to extend the interaction range between qubits beyond the short-range spin−spin interactions commonly employed in solid-state quantum systems.[14,15]

In a recent study, we demonstrated that DAPs in bulk SiC can exhibit optically reconfigurable electric dipole moments, enabling controllable, coherent coupling of ∼ 100 MHz over distances well exceeding 10 nm.[13] First-principles calculations revealed that certain DAPs, particularly Al−N pairs, display well-defined optical transitions and weak electron−phonon coupling, making them attractive for scalable quantum networks. However, practical implementations of DAP-based architectures necessitate further understanding of how these pairs behave in realistic environments, particularly near semiconductor surfaces, and of how surface effects may influence defects' stability and optical properties.[16]

The presence of surfaces introduces both challenges and opportunities for defect-based quantum systems. The local environment of defects at surfaces, including geometric reconstructions, charge trapping, and surface states, may affect their energy levels and optical properties.[17,18] Understanding these effects is crucial for integrating DAPs into practical quantum devices, as many proposed architectures will require near-surface operation, in the presence of interfaces with photonic structures or nanofabricated electrodes.

In this work, we investigate how SiC surfaces influence the electronic and optical properties of near-surface dipoles formed by spatially separated charge centers. We consider two classes of systems: donor−acceptor pairs (DAPs) formed by substitutional impurities, and surface-defect pairs (SDPs), in which a near-surface defect couples to a surface-localized state. While distinct in physical origin, both mechanisms generate optically controllable dipole moments whose orientation and emission characteristics are highly influenced by surface proximity and termination. By analyzing these two cases side by side, we aim to clarify how different defect configurations can be engineered for optically active and directionally aligned dipoles near semiconductor surfaces.

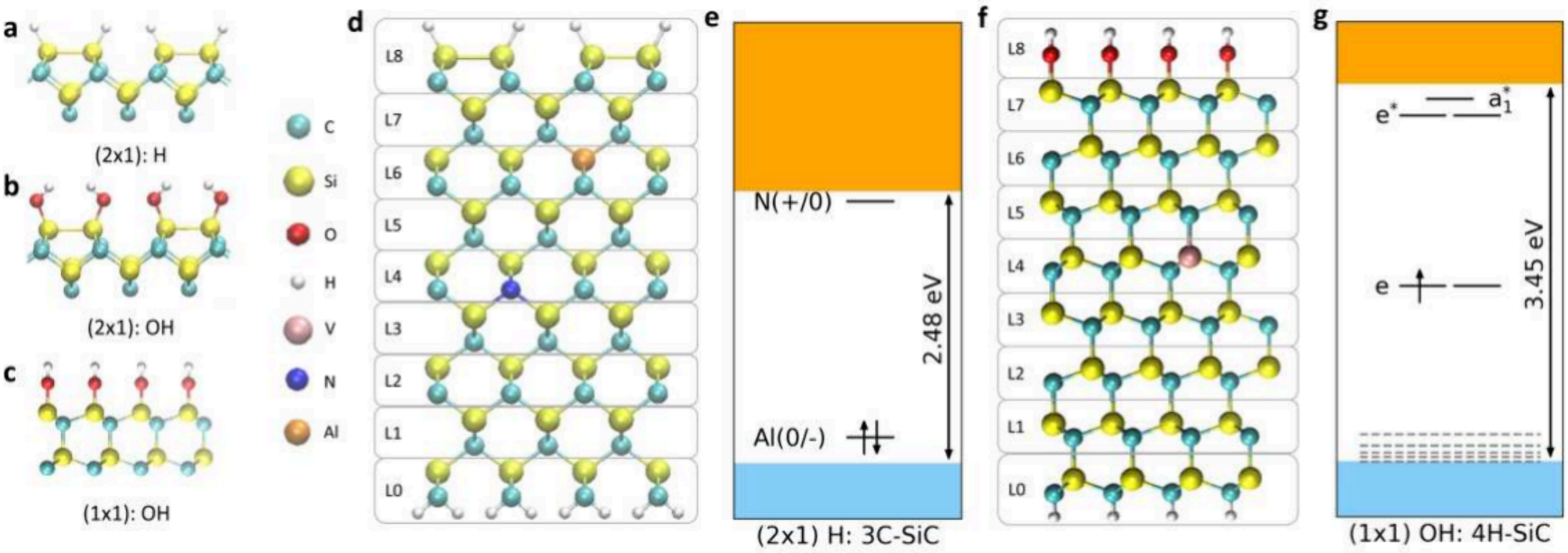

**Figure 1.** Atomistic structures of the Si-rich (001) surfaces of 3C-SiC with (a) a (2 × 1) surface reconstruction terminated with H atoms, (b) a (2 × 1) surface reconstruction terminated with OH groups, and (c) the (0001) surface of 4H-SiC with a (1 × 1) reconstruction terminated with OH groups. (d) Side view of the 3C-SiC slab with the (2 × 1) H-termination, where the bottom C-rich surface is passivated with H atoms. The slab consists of 9 Si−C bilayers, labeled from L0 to L8. The bottom two bilayers, noted as L0 and L1, are kept fixed during geometric optimizations. The top layer L8 is reconstructed, L4 is the middle of the slab, L2 is the closest layer to the fixed bilayers at the bottom, and L7 is the closest bilayer to the surface. Substitutional $Al_{Si}$ and $N_C$ defects are shown in the L4 and L6 layers, respectively. (e) The electronic structure of the slab model is shown in (d). The blue and orange regions indicate the valence and conduction bands. (f) Side view of the 4H-SiC slab with the (1 × 1) OH-termination that includes a $V_{Si}$ defect located in the L4 layer. (g) The electronic structure of the slab model is shown in (f). The gray dashed lines indicate the occupied surface states introduced by the surface termination; the defect levels introduced by $V_{Si}$ are also highlighted.

Our study first extends the investigation of DAPs to hydrogen (H)- and hydroxyl (OH)-terminated surfaces of 3C-SiC and 4H-SiC. Using density functional theory (DFT) calculations, we analyze how donor and acceptor levels are perturbed near the surface, and compute charge transition levels, zero-phonon lines (ZPLs), and electron−phonon coupling parameters to assess their suitability for quantum applications. We then introduce and explore the SDP configuration, wherein an electron is transferred between a near-surface defect and an occupied surface state, forming a dipole aligned perpendicular to the surface. Although this mechanism differs from conventional donor−acceptor binding, its optical signature and dipole strength are comparable. In particular, we examine SDPs formed between substitutional vanadium ($V$) and OH-terminated surfaces—a configuration that could enable optically switchable dipoles in planar, scalable quantum architectures. Similar surface-defect interactions may already have been observed in hydrogen-terminated nanodiamonds, as suggested by recent experiments from Pasternak et al.[19] Taken together, our findings provide a unified framework for designing surface-integrated dipoles based on both impurity pairs and defect−surface interactions, with implications for hybrid quantum-classical interfaces and photonic circuit integration.

Silicon carbide exists in several different polytypes, among which 3C-SiC and 4H-SiC have been extensively studied, both experimentally and theoretically with ab initio calculations.[20] Due to structural differences in the stacking order of SiC bilayers, there are significant differences in bandgap energies, electron mobilities and surface reactivities between SiC polytypes.[21]

SiC surfaces may be C- and Si-terminated, with the latter being the most stable, and exhibit varied reconstructions. To reduce the surface reactivity of Si-terminated SiC surfaces and prevent immediate oxidation, it has been common procedure to hydrogenate them.[17,22] However, despite initial passivation, these hydrogen-terminated surfaces may exhibit limited long-term stability in air, forming native oxide layers over time.[23] Hydroxyl (OH)-terminated Si-rich surfaces of SiC have shown improved chemical stability compared to hydrogen terminated ones.[24] However, they exhibit occupied surface states in the bandgap of the material.[25] Here we study both H and OH-terminated surfaces, recognizing that a complete understanding and control of their long-term stability in air will need to be achieved to optimize their integration in nanofabricated structures for quantum information science.

We consider the atomistic models presented in Figure 1 and we study three different surface reconstructions of two different polytypes of SiC. Specifically, we consider the Si-rich (001) surface of 3C-SiC and we study its (2 × 1) reconstruction terminated either with H atoms or OH groups. We also study the Si-rich (0001) surface of the 4H-SiC polytype, with a (1 × 1) reconstruction terminated with OH groups. We use periodic slab models consisting of 9 bilayers, with 20 Å vacuum between periodic images. We saturate the bottom C-rich surface with H atoms (see Figure 1). The (2 × 1) H-terminated and OH-terminated slabs of 3C-SiC contain 672 and 704 atoms respectively, whereas the slab with the (1 × 1) OH-terminated surface of 4H-SiC has 684 atoms.

An important factor determining whether a surface termination is viable as a platform for quantum technology applications is its electron affinity,[18] defined as the energy difference between the vacuum level and the conduction band minimum (CBM). A positive electron affinity is desirable, as it prevents the loss of excited electrons into the vacuum.[18] Two of the surface terminations considered here, the (2 × 1) H- and (2 × 1) OH-terminated surfaces of 3C-SiC, have previously been reported to have positive electron affinity.[17]

**Table 1. Charge Transition Levels within Bulk and Slab Models**

| Functional | $N_C(+/0)$ (Slab) | $N_C(+/0)$ (Bulk) | $Al_{Si}(0/-)$ (Slab) | $Al_{Si}(0/-)$ (Bulk) | Bandgap (Slab) |
|---|---|---|---|---|---|
| PBE | 0.04 | 0.11 | 0.09 | 0.11 | 1.31 |
| HSE | 0.12 | 0.15 | 0.13 | 0.16 | 2.17 |
| DDH | 0.25 | 0.27 | 0.29 | 0.33 | 2.25 |
| Expt. | - | 0.05−0.07[a] | - | 0.24−0.26[b] | - |

Charge transition levels (CTL) of $N_C$ and $Al_{Si}$ in bulk 3C-SiC and near the (2 × 1) Hydrogen terminated Si-rich surface of 3C-SiC (Slab), computed with three different functionals (PBE,[32] HSE,[33] and DDH[34]), extrapolated to the dilute limit and infinite slab thickness, respectively. The bandgap values are also extrapolated and the slab and bulk extrapolations yield the same number within 0.1 eV. CTLs for $N_C$ and $Al_{Si}$ are given relative to the conduction band minimum and valence band maximum, respectively. Experimental values (Expt.) measured for 3C-SiC at room temperature are also reported. All values are in eV. The bandgap computed in a 9-layered slab is ∼0.1 eV larger than the extrapolated value for all functionals. The CTLs computed in a 9-layered slab are ∼0.13 eV higher (*i.e.,* defects are deeper) than extrapolated values for both defects. [a]Refs. 28−30. [b]Refs. 29−31.

We find that the (1 × 1) OH-terminated surface of 4H-SiC has a positive electron affinity as well (see SI for details).

The electronic structures of the surfaces are summarized in Figure 1. As reported in refs. 17, 25 there are no surface states introduced into the bandgap of the (2 × 1) H-terminated surface, while the (2 × 1) OH-terminated substrate exhibits occupied surface states in the bandgap immediately above the valence band maximum (VBM). Moreover, we find that the (1 × 1) reconstruction of the OH-terminated surface of 4H-SiC exhibits an electronic structure qualitatively similar to that of the (2 × 1) OH-terminated surface of 3C-SiC, with occupied surface states above the valence band. We note that the H- and OH- terminations considered here are intentionally chosen as idealized, uniform coverages to represent diverse surface electrostatic environments: the H-termination yields a clean bandgap with no in-gap states, while the OH-termination introduces occupied surface states just above the valence band. These idealized models allow us to systematically study how DAP and SDP properties are influenced by different electrostatic environments near the surface. While the terminations chosen here may not be thermodynamically stable under all experimental preparation conditions, they are useful for revealing robust trends. Prior studies have shown that mixed H/OH terminations can offer improved chemical stability and more effective suppression of surface states.[18,26] Additionally, epitaxial passivation approaches such as thin AlN layers[27] represent promising alternatives for experimental implementations. The study of detailed mixed or epitaxial passivation schemes is beyond the scope of this work but presents an important direction for future investigations.

The differences in electronic structure between the H- and OH-terminated surfaces present different opportunities for their use. The absence of gap states in the former avoids possible interferences with optical transitions between defect levels. This property makes the system suitable for operations of DAPs near surfaces for quantum sensing and enhanced control through fabricated cavities and waveguides. On the other hand, the introduction of occupied surface states near the OH-terminated surfaces present the opportunity to pair them with those of another defect close to the surface and to create a combined surface-defect pair (SDP). While in DAPs the electron is bound to the donor and the hole to the acceptor, in SDPs, the hole is a surface state near the VBM and the excited electron is localized on the defect.

In the next few paragraphs we analyze DAPs near silicon carbide surfaces. We consider Al−N pairs, found in our previous study[13] to be promising ones, given their weak electron−phonon coupling and efficient emission into the ZPL. We position $N_C$ donors in the middle bilayer of the slab (L4 in Figure 1) and we vary the position of $Al_{Si}$ acceptors, thus varying the donor−acceptor pair distance ($R_m$) and the relative orientation with respect to the surface.

In Table 1 we present a summary of the charge transition levels (CTLs), computed in the bulk and near the surfaces, for $N_C$ donors and $Al_{Si}$ acceptors. After extrapolating the results to the dilute limit for the bulk and to an infinitely thick slab (see SI), we find that the CTLs computed in the slab are only marginally shallower than those in the bulk. The difference is such that even at a moderate temperature the bulk and surface values of the CTLs are to be considered the same. However, when using a 9-layered slab, we find an increase of ∼ 0.16−0.18 eV in the computed bandgap and a variation of ∼ 0.13 eV in the CTLs. These differences are due to quantum confinement.[35,36] Nevertheless, we carry out our calculations for DAPs with a slab thickness of 9 bilayers for reasons of computational efficiency, keeping in mind the small error relative to bulk values due to finite size effects, when we discuss trends.

As well-known, the computed PBE value of the bulk bandgap in 3C-SiC is underestimated (1.40 eV), compared to the experimental value of 2.36 eV.[21] Both the HSE and DDH functionals yield reasonable values, at 2.27 and 2.35 eV respectively. The HSE functional predicts $Al_{Si}$ and $N_C$ to be shallower defects, closer to their respective band edges, compared to the DDH functional. For $N_C$, which has a (+/0) charge transition level ∼ 50−70 meV below the CBM,[28−30] measured at room temperature, the HSE and DDH predictions in bulk are 0.15 and 0.27 eV respectively. In the slab, these predictions are 0.12 and 0.25 eV respectively. Similarly for $Al_{Si}$, which is a shallow acceptor with the (0/−) transition level 250 meV above the VBM[29−31] at room temperature, the values computed in bulk using HSE and DDH are 0.16 and 0.33 eV respectively. In the slab, these values are 0.13 and 0.29 eV respectively.

While neither functional yields a perfect agreement with experiments for the charge transition levels of both $Al_{Si}$ and $N_C$, the performance of HSE appears to be slightly better. However, note that the differences between the results with the HSE and DDH functionals (∼100 meV) are on the order of the zero point renormalization of the bandgap of SiC.[37,38] Hence it is not straightforward to determine whether the performance of one functional is definitely superior.

Next we determine the zero-phonon lines (ZPL) of Al−N DAPs near (2 × 1) H-terminated SiC surfaces and their respective photoluminescence (PL) spectra. We compute ZPL values using the HSE functional, and determine the Debye−

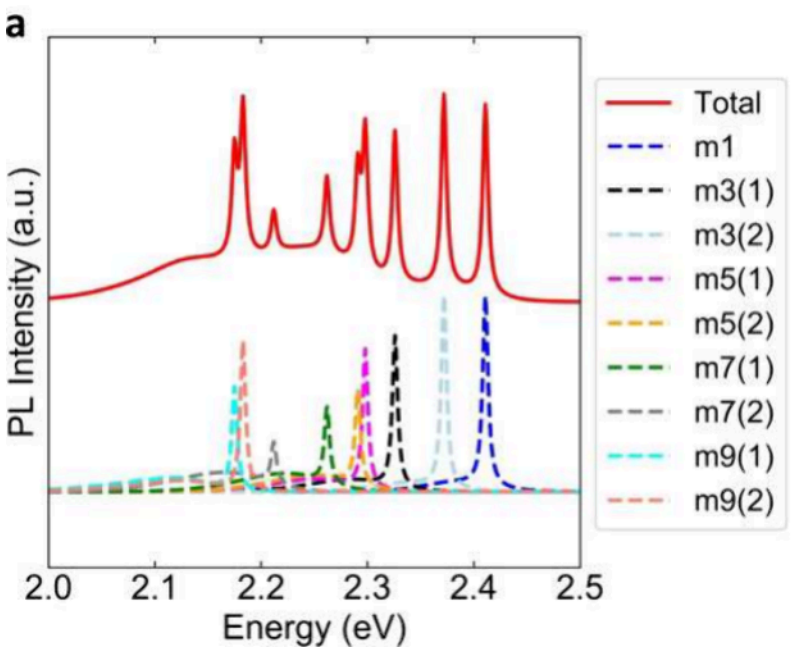

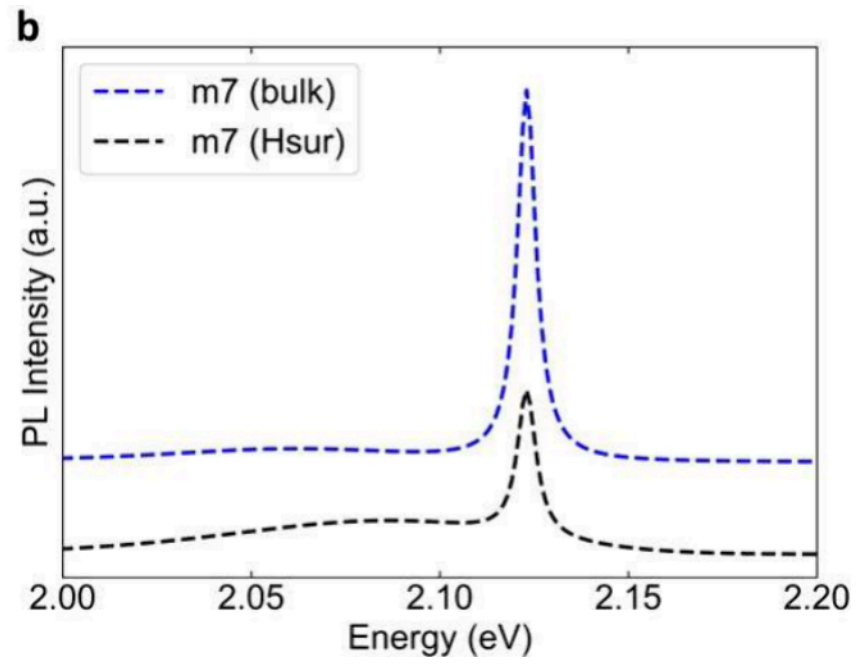

**Figure 2.** (a) Cumulative photoluminescence (PL) spectra (red line) of the first 9 shells (m1−m9) of Al−N DAPs near a (2 × 1) H-terminated Si-rich surface of 3C-SiC. The individual contributions to the PL spectra are shown below the cumulative one. (b) A comparison of the PL spectra coming from the m7 shell of Al−N pairs in bulk (blue dashed line) and near a (2 × 1) H-terminated surface (Hsur) of 3C-SiC (black dashed line). The zero-phonon line (ZPL) values are computed using the HSE functional, and the phonon sidebands are computed using a scaling factor based on the PBE functional (see text). The near-surface ZPL value is aligned to match the ZPL value computed in the bulk for comparison (see text).

Waller factors (DWF) and PL spectra using a one-dimensional model.[39−43] We approximate the effective phonon frequencies in the ground and excited states by using a scaling factor based on the PBE results following the same method outlined in ref.[42] This scaling factor is evaluated from the ratio of frequencies of the bulk system's optical phonons computed at different levels of theory (PBE and HSE). Using a 9-layered slab, we find a 0.2 to 0.3 eV increase in ZPL energy near the SiC surface compared to the bulk values. Given that the difference between computed Franck−Condon shifts in the bulk and in the 9-layered slab is only ∼ 10−15 meV, and given the extrapolated results presented in Table 1, we conclude that the main reason for such an increase of the surface ZPL comes again from finite size effects. Therefore, we conclude that surface and bulk ZPLs do not show any notable change, especially at finite temperature.

We find that ZPL energies from different shell numbers decrease as a function of the pair distance $R_m$, due to the weakening of the Coulomb interaction with distance.[44−48] In all cases we observe that DAPs near surfaces exhibit larger mass-weighted displacement ($\Delta Q$) values compared to their bulk counterparts. This results in a more pronounced Huang−Rhys (HR) coupling[49] and hence leads to a decrease in the DWF. We also find that DAPs with the same pair distance $R_m$ but with different orientations relative to the surface, may exhibit different $\Delta Q$ values. In most cases, a considerable increase in $\Delta Q$, relative to the bulk, is associated with the $Al_{Si}$ acceptor being placed in close proximity to the surface (in the L6 and L7 layers of the slab; see Figure 1). When the $Al_{Si}$ acceptor is placed in the lower layers (L3-L5), the $\Delta Q$ value is lower. The dominant contributions to the increased $\Delta Q$ arise from surface-proximal Si atoms, whose displacements drive the overall reconfiguration. We expect this tendency to be common but not universal, as the magnitude of $\Delta Q$ shifts is sensitive to defect species, orientation, depth, and the specific surface termination.

The $\Delta Q$ values for Al−N pairs near the surface range from 0.32 to 0.88 amu$^{1/2}$Å compared to bulk values of 0.1−0.25 amu$^{1/2}$Å. Similarly, the corresponding HR factors for Al−N pairs near the surface range from 0.35 to 1.9, compared to their bulk counterparts of 0.1 to 0.4. In spite of this three to 4-fold increase, the absolute value of the $\Delta Q$ for near surface pairs remains relatively low (e.g., compared to that of the NV-center in diamond, with HR factor of ∼ 3.6[50]), and should not hinder quantum technology applications. In Figure 2 we show the cumulative PL spectra originating from multiple shells (m1-m9) of Al−N pairs near the (2 × 1) H-terminated surface of 3C-SiC as well as a comparison of the PL spectra from a single shell (m7) in the bulk and near the (2 × 1) H-terminated surface. Note that the contribution of each shell to the PL spectra is normalized, thus highlighting the spectral shape rather than the absolute intensity. While the absolute PL intensity expected to decrease with increasing DAP separation due to the diminishing transition dipole moment, this scaling is not reflected in the normalized spectra.

Despite the increased $\Delta Q$ and HR factors, near surface Al−N pairs exhibit a sharp, narrow emission into the ZPL. We also observe a broadband shoulder at low energies, originating from the sum of all emissions into the phonon sidebands, a phenomenon widely observed with DAPs in semiconductors.[44,45] Experimentally these broad bands are pronounced and found for high shell numbers (usually after 30th or 40th shell[44−48]); in our computational model we just observe an onset for shell numbers as low as m9. The comparison of PL spectra for a single shell (m7) in bulk and near-surface (Figure 2b) shows a decreased emission into the ZPL and a more pronounced phonon sideband for near-surface pairs.

In sum, the main difference found for near-surface DAPs compared to their bulk counterpart amounts to slightly higher $\Delta Q$ values, and consequently, an increased electron−phonon coupling. However, this variation does not significantly affect the electronic structure or the physical properties of the DAP system in detrimental ways for applications in quantum science platforms. Importantly, both in bulk and near the surface, DAPs exhibit large electric dipole moments which render them highly sensitive to electric fields through a Stark shift, and thus useful for quantum sensing applications. In our previous work on DAPs in bulk diamond and SiC,[13] we computed Stark shifts and demonstrated that electric fields of ∼ 0.5 MV/cm produce ZPL shifts of several THz. Since Stark shifts scale with dipole moment and are further enhanced by reduced dielectric screening near surfaces, we expect comparable or greater sensitivity for near-surface systems investigated here. In addition, a possible placement of DAPs directly underneath nanofabricated structures such as waveguides and cavities would be beneficial to enhance the optical properties of the nanostructures.

Having examined how near-surface electrostatic interactions influence donor−acceptor pairs formed by substitutional impurities, we now turn to a distinct but related system: the formation of dipoles involving a near-surface defect (vanadium) and occupied surface states arising from the −OH surface termination. While the physical origin of the dipole differs - donor−acceptor binding in the former case, surface-induced charge transfer in the latter—both systems exhibit tunable, spatially separated charges and optically addressable dipole moments. To explore the surface-defect pair (SDP) system in detail, we construct a model with a substitutional vanadium impurity ($V^{4+}$, hereafter denoted simply as $V$) in the presence of occupied surface states arising on the (2 × 1) OH-terminated 3C-SiC surface and the (1 × 1) OH-terminated 4H-SiC surface. In this system, an electron is transferred from the highest occupied surface state into a $V$ defect orbital, leaving behind a hole localized at the surface. The resulting excitonic wave function of the surface-vanadium pair is shown in Figure 3, and yields an electric dipole moment aligned perpendicular to the surface and parallel to other SDPs, as we verify below.

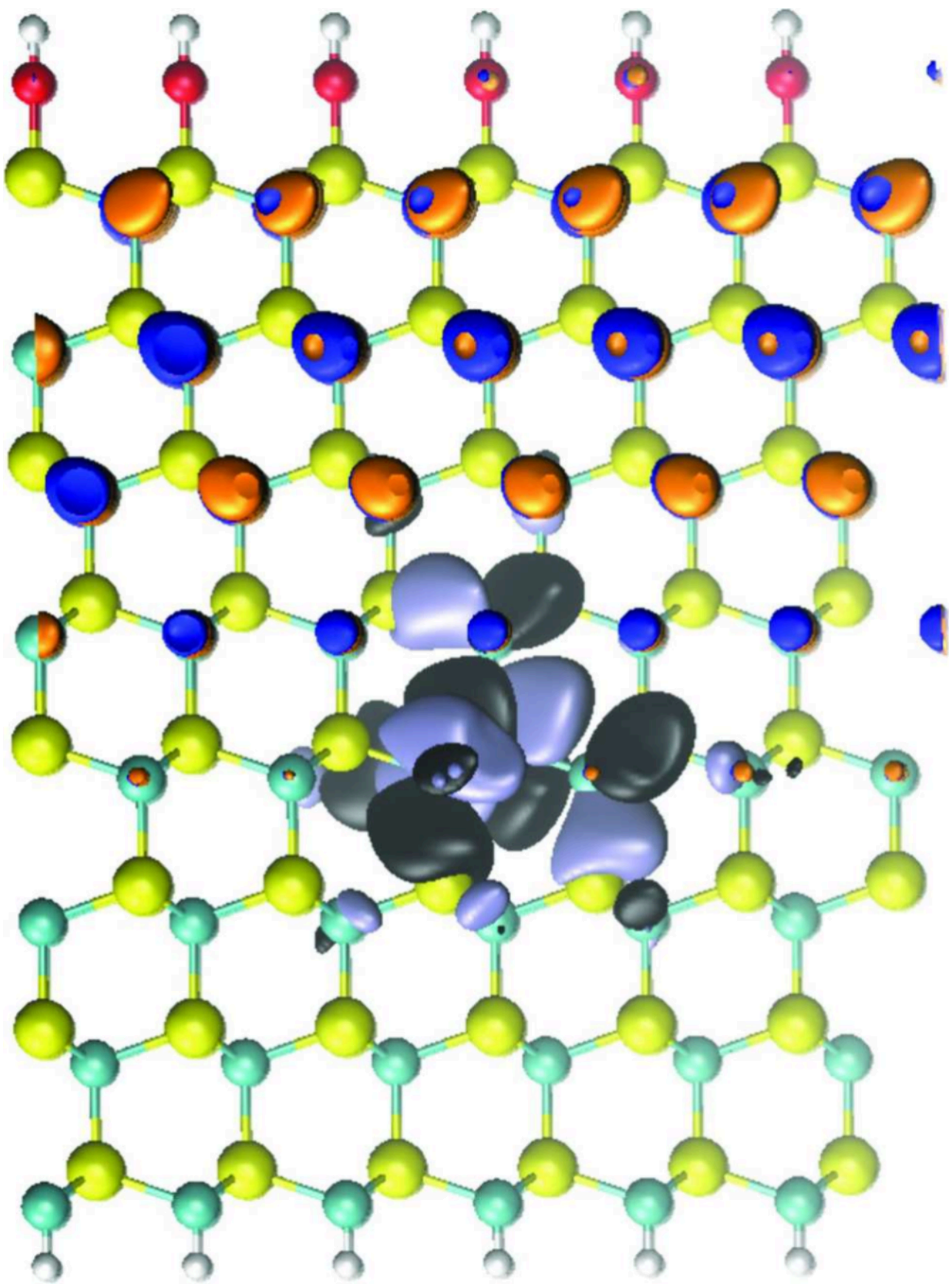

**Figure 3.** Snapshot of the surface-Vanadium pair's excitonic wave function. Isosurfaces of the square modulus of Kohn−Sham wave functions for the surface hole state (blue-orange) and for the Vanadium $e$ state (black-ice blue) are shown for the 4H-SiC slab with a (1 × 1) OH-terminated surface. The isosurface value is set at 0.001 e/Å$^3$. The blue-orange and black-ice blue colors represent the sign (±) of the orbitals respectively.

While the specific surface states modeled here originate from a uniform OH-termination, the broader mechanism highlighted in our work - namely, charge transfer between a surface-localized state and a near-surface defect - is not limited to this particular configuration. Indeed, similar dipole-forming interactions may occur via interface traps, chemical adsorbates, or bound surface charges in more complex or mixed termination schemes. Hence, the SDP model is intended not as a proposal for a specific surface recipe, but as a conceptual framework to understand the potential and pitfalls of using proximal surface states to control dipole orientations. Such surface-induced charge dynamics may also be responsible for photoinstabilities observed in experiments, such as blinking or bleaching,[18,51] reinforcing the need for robust surface engineering strategies to stabilize emitter charge states. Our results point to both opportunity and caution in this regime, highlighting routes to controllable dipole alignment as well as identifying possible pathways for optical instability.

The electronic structure of Vanadium in bulk SiC is well studied.[52−56] In Figure 1 we show the electronic structure of a near surface $V$ atom which is similar to that in the bulk; hence the presence of occupied surface states just above the VBM does not substantially affect the $V$ electronic structure. Vanadium is stable as a neutral defect in a wide range of doping levels[57] and has a spin doublet ground state.[54−56] It introduces two degenerate $e$ states, of which only one is singly occupied in the band gap. These $e$ states are energetically much higher than those of the occupied surface states introduced by the OH surface termination. The excitation from the occupied surface states into one of the unoccupied $e$ states corresponds to the lowest possible optical excitation.

The shortcomings of the PBE and HSE functionals in describing the transition metal $d$ orbitals have previously been documented in ref.[52,57] Here we use the DDH functional to describe surface-Vanadium pairs, which yields (+/0) and (0/−) CTLs of $V$ more accurately, compared to HSE, and without the need for a non-Koopman's correction scheme.[52] Experimentally, the (+/0) and (0/−) transitions are found at 1.6−1.7 and 2.2−2.3 eV above the VBM[53,58] respectively. In our study, we find the (+/0) and (0/−) levels to be 1.18(1.32) and 1.98(2.09) using the HSE (DDH) functional in bulk 4H-SiC. We find that these CTLs in bulk 3C-SiC and near the surface in our slabs are all within ≤ 0.1 eV of the above-mentioned levels in bulk 4H-SiC.

Using the DDH functional, we find that the ZPL energies of Vanadium SDPs in 3C-SiC are at ∼ 1.38 eV, an energy higher than that required to ionize the neutral $V$ and excite an electron from the singly occupied $e$ state into the conduction band. This ionization energy is the difference between the energy of the CBM and the (+/0) charge transition level, and is ∼ 1.0 eV in our calculations, since the bandgap at the DDH level is 2.35 eV. Therefore the Vanadium SDPs would be unstable in 3C-SiC, i.e. subject to ionization. However, we find that using 4H-SiC, instead of 3C, provides an alternative to stabilizing Vanadium SDPs, due to the larger band gap of the hexagonal polytype. The band gap of 4H-SiC is accurately reproduced by the DDH functional (3.23 eV), and the ZPL of Vanadium SDPs near the 4H-SiC surface are found at about 1.47 eV. As mentioned above, the relevant CTLs, (+/0) and (0/−) are 1.32 and 2.09 eV respectively. Hence the ionization thresholds are at CBM - (+/0) ≃ 1.9 eV and (0/−) - VBM ≃ 2.1 eV, both of which are above the ZPL energy, even if we include a considerable underestimation (0.2−0.3 eV) of the (+/0) transition level due to finite size effects and/or inaccuracies of the functional used here. These results indicate that Vanadium SDPs near (1 × 1) OH-terminated surfaces of 4H-SiC are optically viable.

We next examine the alignment of the SDPs electric dipole moments perpendicular to the surface, and parallel to other neighboring SDPs. We compute the electric dipole moment according to the Modern Theory of Polarization[59] using maximally localized Wannier functions (MLWF) and the Wannier90 package.[60] In Figure 4 we show how the electric dipole moment of Vanadium SDPs change as a function of the Vanadium's distance from the surface. We see that the transverse (x and y) components of the SDP remain relatively constant while the longitudinal (z) component scales linearly as the Vanadium resides in different bilayers (L3, L4 etc.) and its distance from the surface (L8) is increased. The fluctuations in the transverse components are on the order of those found in our previous study in the bulk.[13] We also note that the $\Delta Q$ value of Vanadium SDPs near $(1 \times 1)$ OH-terminated 4H-SiC surfaces is $\sim$ 0.606 amu$^{1/2}$Å, which translates to an HR coupling value of $\sim$ 2.6. Therefore, the SDPs have a weaker emission into the ZPL than the Al–N DAPs discussed in the previous section; however note that their emission is still superior to that of the NV- center in diamond. Combining our results that show that the dipole moment is indeed aligned perpendicular to the surface and an acceptable level of electron–phonon coupling, Vanadium SDPs present several advantages: It offers an opportunity to use SDPs in quantum science platforms by leveraging their strong dipolar interactions. The natural alignment of dipoles makes their interaction stronger than in the bulk where dipoles of DAPs are randomly oriented. Furthermore, SDPs may serve as ideal candidates for an array-like quantum sensor platform that can be interfaced with other nanofabricated structures at the SiC surface.

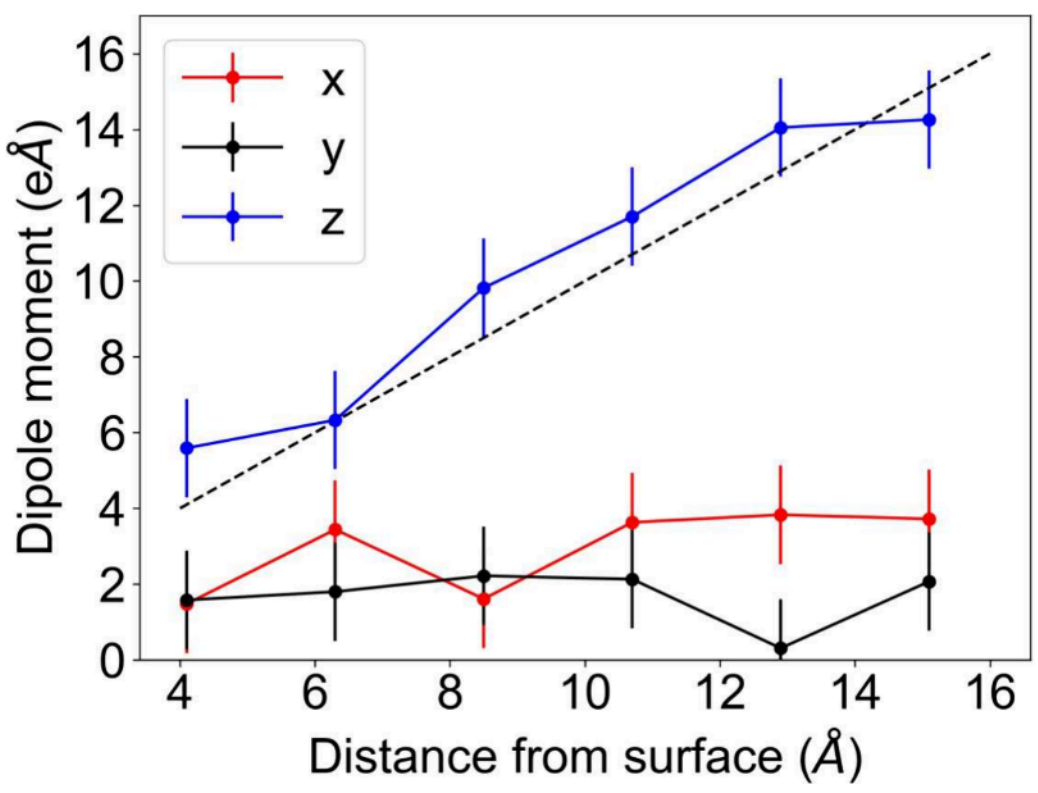

**Figure 4.** Transverse (x and y) and longitudinal (z) components of the electric dipole moment of surface-defect pairs formed between Vanadium and the $(1 \times 1)$ OH-terminated surface of 4H-SiC. The transverse components remain almost unchanged and the longitudinal component scales linearly with the distance of Vanadium from the surface. The black dashed line is a $y = x$ line meant as a guide to the eye.

Finally, we compute the stimulated emission ($\sigma_{st}$) and photoionization ($\sigma_{ion}$) cross sections of Vanadium SDPs to investigate different optical control mechanisms. They are defined as

$$\sigma_{st}(\hbar\omega, T) = \frac{4\pi^2\alpha}{n}\hbar\omega|\mathbf{r}_s|^2 A(\hbar\omega - E_{ZPL}, T)$$

$$\sigma_{ion}(\hbar\omega, T) = \frac{4\pi^2\alpha}{n}\hbar\omega \sum_j |\mathbf{r}_j|^2 A(\hbar\omega - E_j, T) \quad (1)$$

where $\alpha$ is the fine structure constant, $n$ is the refractive index, $\mathbf{r}_j$ and $E_j$ are the transition dipole moment matrix elements and the energies of the transitions respectively. $A(\hbar\omega)$ is the electron–phonon spectral function. Following refs. 39–43 we use a one-dimensional approach to obtain the electron–phonon spectral function $A(\hbar\omega)$ and compute the photoionization and stimulated emission cross sections. The transition dipole moment matrix elements are computed with the WEST code.[61,62] In our calculation of the stimulated emission, we use the transition dipole matrix element $|\mathbf{r}_s|$ that corresponds to the ZPL transition; for the photoionization calculation, we include transition matrix elements corresponding to ionizations at multiple energies. By considering a generalized incident laser light, we determine the transition dipole moment magnitude as a linear combination of dot products between the computed dipole moment and the two polarization vectors of the incident laser. Further details of this derivation are given in the SI. We show the computed stimulated emission cross sections of Vanadium SDPs in Figure 5. The left (right) column of the figure presents the case of the incident laser parallel (perpendicular) to the $z$ axis. We note that only the transition dipole moment between the hole state at the surface and the occupied $e$ state in Vanadium is contributing to the stimulated emission. As expected, this contribution is the strongest when the Vanadium atom resides close to the surface (L6, L7 layers). The weakening of the contribution as $V$ is placed further from the surface is significant: moving the Vanadium atom by even a few Ångstroms from the surface may result in more than a 10-fold decrease in the magnitude of the transition.

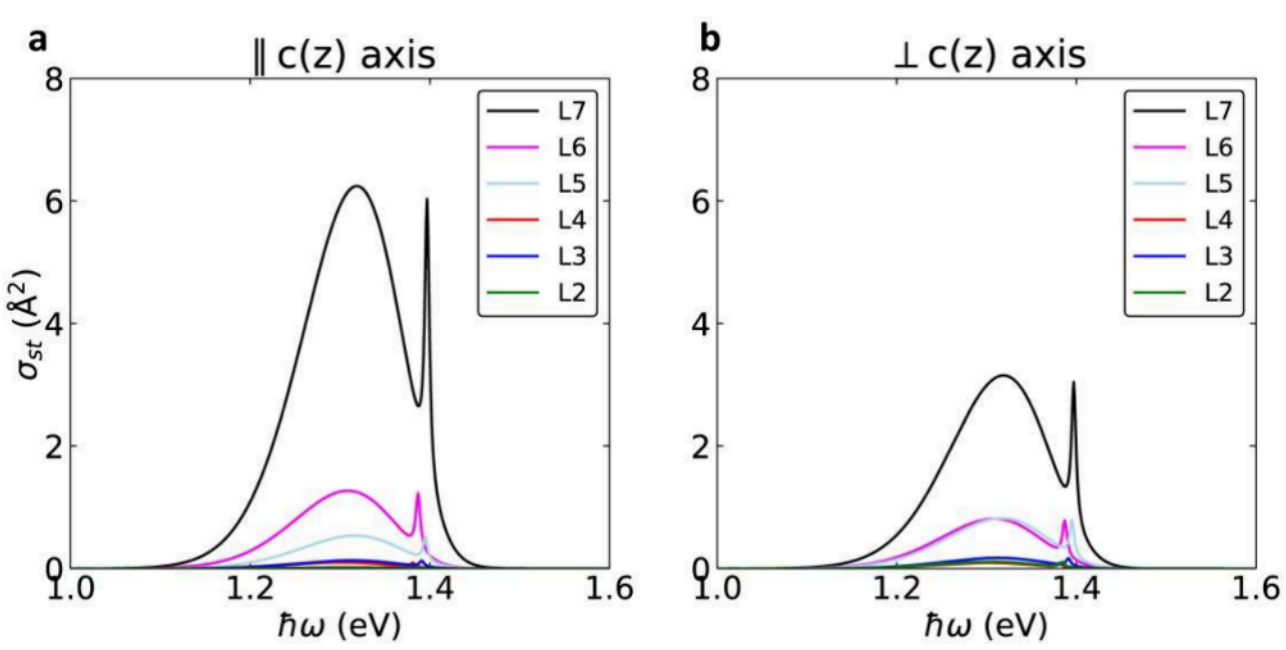

**Figure 5.** Stimulated emission cross sections ($\sigma_{st} \times 10^{-3}$) for Vanadium surface-defect pairs near OH-terminated SiC surfaces for incident exciting laser (a) parallel to the $z$ axis and (b) perpendicular to the $z$ axis. Stimulated emission cross sections from slabs containing a $V$ defect at different bilayers (see Figure 1) from L2 to L7 are shown together in both panels. We use a one-dimensional approach to compute the electron–phonon spectral function $A(\hbar\omega)$ and compute the transition dipole matrix elements using the WEST code[61,62] (see text).

The line shapes of the photoionization cross section shown in Figure S1 are determined by different transition matrix elements, depending on the distance of $V$ from the surface. For example, when the Vanadium is in close proximity to the surface (L6, L7 layers), the transition matrix element corresponding to the ZPL is dominant (as it is in the stimulated emission case), and gives rise to the low energy peak at 1.50 eV in Figure S1. Instead, when $V$ is further from the surface (e.g., in L3-L5), the photoionization process occurs at higher energies. We note that depending on the incident laser light angle with the surface, the contributions to the sum entering eq 1 vary.

It is thus interesting to investigate how the cross section depends on the polarization of the incident laser and we do so using the following equation:

$$|\mathbf{r}_j|^2 = |\mathbf{e}\cdot\mathbf{r}_j|^2 = |\cos\delta\mathbf{e}_{\parallel}\cdot\mathbf{r}_j + \sin\delta\mathbf{e}_{\perp}\cdot\mathbf{r}_j|^2 \quad (2)$$

where $\mathbf{r}_j$ are transition dipole moment matrix elements, $\delta$ is the polarization angle of a linearly polarized incident laser light, and $\mathbf{e}_{\parallel}$ and $\mathbf{e}_{\perp}$ are the two polarization vectors for the incoming laser light. In eq 2, we expand the transition matrix element as a linear combination of dot products, each of which is a projection onto the polarization vectors of the incident laser (See SI for details). By using linearly polarized light, the magnitudes of both the ionization and the stimulated emission cross sections can be modulated. This allows us to determine from first-principles calculations the specific polarization angles $\delta$ at which the optical processes of photoionization and stimulated emission can either be enhanced or suppressed. The locations of the peak and trough and the magnitude of the change in cross sections as a function of $\delta$ can be expressed in an analytical form (see SI). Surprisingly, we find that the difference in magnitude between the peak and the trough can be one hundred-fold, offering 2 orders of magnitude tuning range for enhancement or suppression of these processes, thus offering a unique tool to control the SDPs.

We have shown that the combination of long-range interactions and optical addressability makes DAPs and SDPs near SiC surfaces compelling candidates for quantum technologies. Specifically, optically addressable dipole moments are critical for controlling long-range dipole−dipole interactions in solid-state quantum platforms. By leveraging the inherent properties of SiC, including its compatibility with existing semiconductor processing techniques and its high thermal stability, the defects studied here—Al−N DAPs and surface-Vanadium SDPs—could serve as foundational elements for robust, scalable quantum platforms.

Interestingly, we have found that DAPs near H-terminated SiC surfaces maintain most of their favorable properties relative to their bulk counterparts, making them viable candidates for near-surface quantum networks. On the other hand, SDPs formed with Vanadium defects at OH-terminated 4H-SiC surfaces exhibit naturally aligned dipole moments perpendicular to the surface, which enhances their dipolar interaction strength significantly compared to the bulk. This property paves the way for engineering 2D arrays of quantum defects that can be exploited for scalable quantum sensing and quantum communication applications.

Another critical insight of our work is the role played by electron−phonon coupling in shaping the optical properties of near-surface defects. Our results suggest that while near-surface placement can slightly increase the electron−phonon interaction, the overall optical emission characteristics of DAPs remain favorable. Importantly, the photostability of SDPs is strongly dependent on their charge transition levels relative to the conduction and valence band edges. In 3C-SiC, we find that certain SDP configurations are prone to unwanted photoionization, whereas the wider bandgap of 4H-SiC provides a more stable environment for optical excitation and dipole interaction control.

Moreover, we have explored the polarization dependence of stimulated emission and photoionization under linearly polarized light and shown that it is possible, entirely from first-principles, to determine the optimal polarization angle to enhance or suppress such optical processes in DAPs and SDPs. We note that the determination of polarization angles is general and can be applied to many other defects of interest. One example could be determining the optimal polarization angle during spin to charge conversion readout experiments involving NV centers in diamond.[63] While our current study focuses on linearly polarized excitation for simplicity and experimental relevance, the extension of our approach to arbitrary polarization bases would be a valuable direction in the future.

Future work should focus on validating our theoretical predictions through experimental studies, including controlled fabrication of near-surface defects and direct measurements of their optical properties. Among the challenges to be addressed is the air-stability of H-terminated surfaces, which is only of the orders of tens of hours and needs to be greatly improved. In addition, advancements in atomic-scale defect placement techniques, such as deterministic ion implantation and in situ processing during epitaxial growth, will be essential for realizing practical quantum devices based on near-surface SiC defects. Further, integration with photonic nanostructures and waveguides will be crucial for optimizing light-matter interactions and enhancing the scalability of these systems.

We determine the electronic structure of donor−acceptor pairs (DAPs) and surface-defect pairs (SDPs) in 3C and 4H-SiC using spin-polarized density functional theory (DFT), and the planewave pseudopotential method as implemented in the Quantum Espresso package.[64−66] We use plane wave basis sets with a kinetic energy cutoff of 80 Ry and SG15 Optimized Norm-Conversing Vanderbilt (ONCV) pseudopotentials.[67,68] We carry out lattice optimizations to determine the charge transition levels (CTLs) of $N_C$, $Al_{Si}$, and $V$ using three different functionals, namely PBE,[32] HSE,[33] and DDH.[34] The screening and mixing parameters used for HSE are 0.2 Å$^{-1}$ and 25% respectively. For the DDH functional, we use a mixing parameter $\epsilon_{\infty}^{-1}$ = 0.153. For determining other properties, we use lattice optimizations based on the PBE functional, but compute the electronic structure using HSE and DDH, and mention which one is used in the main text. All lattice optimizations use a force convergence threshold of 10 meV/Å. We sample the Brillouin zone at the $\Gamma$ point in all calculations. The excited states of the defects are investigated using the constrained DFT method.

We determine the electric dipole moments of DAPs and SDPs according to the Modern Theory of Polarization[59] using maximally localized Wannier functions (MLWF) and the Wannier90 package.[60] We compute the ZPL values using both the HSE and DDH functionals, and in the text we mention which one is used. We determine the Debye−Waller factors (DWF) and the phonon sideband of the PL spectra using a one-dimensional model to obtain one effective phonon mode.[39−43] We approximate the effective phonon frequencies in the ground and excited states by using a scaling factor based on the PBE results following the same method outlined in ref.[42] This scaling factor is evaluated from the ratio of frequencies of the bulk system's optical phonons computed at different levels of theory. In our calculations for the photoionization and stimulated emission cross sections, we determine the electron−phonon spectral function $A(\hbar\omega)$ based on this one-dimensional method. We compute the transition dipole matrix elements using the WEST code.[61,62]

For stimulated emission and photoionization calculations the matrix elements are computed in the excited state and ground state respectively.

In our slab calculations, we use 9-bilayer slabs to study three different surface reconstructions of two different polytypes of SiC. We consider the Si-rich (001) surface of 3C-SiC, and study its (2 × 1) reconstruction terminated with either H atoms or OH groups. We also study the Si-rich (0001) surface of 4H-SiC with a (1 × 1) reconstruction terminated with OH groups. The bottom C-rich surface is saturated with H atoms, and 20 Å of vacuum is added between periodic images of the slabs. To remove the artificial electrostatic field introduced by the asymmetry of the slab used in our calculations, we apply a dipole correction by adding a sawtooth potential across the entire supercell.[69] This ensures that the vacuum level remains constant throughout the supercell and that computed electrostatic potentials and Kohn−Sham energy levels are not affected by spurious long-range dipole interactions. The bottom two bilayers of the slab (L0 and L1) are kept fixed during relaxation. The (2 × 1) H-terminated and OH-terminated slabs of 3C-SiC contain 672 and 704 atoms, respectively, while the (1 × 1) OH-terminated 4H-SiC slab contains 684 atoms.

## ■ ASSOCIATED CONTENT

### Supporting Information

The Supporting Information is available free of charge at https://pubs.acs.org/doi/10.1021/acs.jpclett.5c02376.

> Detailed information about how the electron affinity of surfaces are computed, the finite size effects both in bulk and in slab calculations, and the full derivation for how the polarization control of photoionization and stimulated emission cross-sections are achieved (PDF)
>
> Transparent Peer Review report available (PDF)

## ■ AUTHOR INFORMATION

### Corresponding Authors

**Anil Bilgin** − *Pritzker School of Molecular Engineering, The University of Chicago, Chicago, Illinois 60637, United States;* orcid.org/0000-0002-7938-8487; Email: bilgin@uchicago.edu

**Giulia Galli** − *Pritzker School of Molecular Engineering and Department of Chemistry, The University of Chicago, Chicago, Illinois 60637, United States; Center for Molecular Engineering and Materials Science Division, Argonne National Laboratory, Lemont, Illinois 60439, United States;* orcid.org/0000-0002-8001-5290; Email: gagalli@uchicago.edu

### Authors

**Ian N. Hammock** − *Pritzker School of Molecular Engineering, The University of Chicago, Chicago, Illinois 60637, United States*

**Alexander A. High** − *Pritzker School of Molecular Engineering, The University of Chicago, Chicago, Illinois 60637, United States; Center for Molecular Engineering and Materials Science Division, Argonne National Laboratory, Lemont, Illinois 60439, United States*

Complete contact information is available at: https://pubs.acs.org/10.1021/acs.jpclett.5c02376

### Author Contributions

A.B. performed all DFT calculations. I.H. designed the experimental setup and obtained PL spectra from SiC samples. A.H. and G.G designed and supervised the research. A.B. and G.G wrote the manuscript.

### Notes

The authors declare no competing financial interest.

## ■ ACKNOWLEDGMENTS

This work was supported by AFOSR Grant No. FA9550-22-1-0370. We acknowledge support from Boeing through Chicago Quantum Exchange, and the computational materials science center Midwest Integrated Center for Computational Materials (MICCoM). This research used resources of the National Energy Research Scientific Computing Center (NERSC), a DOE Office of Science User Facility supported by the Office of Science of the U.S. Department of Energy, under Contract No. DE-AC02-05CH11231. We acknowledge funding from NSF QLCI for Hybrid Quantum Architectures and Networks (NSF award 2016136). This work also made use of the shared facilities at the University of Chicago Materials Research Science and Engineering Center, supported by the National Science Foundation under Award Number DMR-2011854, as well as the resources provided by the University of Chicago's Research Computing Center.

## ■ REFERENCES

(1) Awschalom, D. D.; Hanson, R.; Wrachtrup, J.; Zhou, B. B. Quantum Technologies with Optically Interfaced Solid-State Spins. *Nat. Photonics* **2018**, *12*, 516−527.

(2) Weber, J. R.; Koehl, W. F.; Varley, J. B.; Janotti, A.; Buckley, B. B.; Van de Walle, C. G.; Awschalom, D. D. Quantum Computing with Defects. *Proc. Natl. Acad. Sci. U. S. A.* **2010**, *107*, 8513−8518.

(3) Bassett, L. C.; Alkauskas, A.; Exarhos, A. L.; Fu, K. C. Quantum Defects by Design. *Nanophotonics* **2019**, *8*, 1867−1888.

(4) Wolfowicz, G.; Heremans, F. J.; Anderson, C. P.; Kanai, S.; Seo, H.; Gali, A.; Galli, G.; Awschalom, D. D. Quantum Guidelines for Solid-State Spin Defects. *Nat. Rev. Mater.* **2021**, *6*, 906−925.

(5) Sipahigil, A.; Evans, R. E.; Sukachev, D. D.; Burek, M. J.; Borregaard, J.; Bhaskar, M. K.; Nguyen, C. T.; Pacheco, J. L.; Atikian, H. A.; Meuwly, C.; Camacho, R. M.; Jelezko, F.; Bielejec, E.; Park, H.; Lončar, M.; Lukin, M. D. An Integrated Diamond Nanophotonics Platform for Quantum-Optical Networks. *Science* **2016**, *354*, 847−850.

(6) Lenzini, F.; Gruhler, N.; Walter, N.; Pernice, W. H. P. Diamond As a Platform for Integrated Quantum Photonics. *Adv. Quantum Technol.* **2018**, *1*, 1800061.

(7) Ruf, M.; Wan, N. H.; Choi, H.; Englund, D.; Hanson, R. Quantum Networks Based on Color Centers in Diamond. *J. Appl. Phys.* **2021**, *130*, 070901.

(8) Yi, A.; Wang, C.; Zhou, L.; Zhu, Y.; Zhang, S.; You, T.; Zhang, J.; Ou, X. Silicon Carbide for Integrated Photonics. *Appl. Phys. Rev.* **2022**, *9*, 031302.

(9) Castelletto, S. Silicon Carbide Single-Photon Sources: Challenges and Prospects. *Mater. Quantum Technol.* **2021**, *1*, 023001.

(10) Majety, S.; Saha, P.; Norman, V. A.; Radulaski, M. Quantum Information Processing with Integrated Silicon Carbide Photonics. *J. Appl. Phys.* **2022**, *131*, 130901.

(11) Cai, H.; Rasmita, A.; Tan, Q.; Lai, J. M.; He, R.; Cai, X.; Zhao, Y.; Chen, D.; Wang, N.; Mu, Z.; Huang, Z.; Zhang, Z.; Eng, J. J. H.; Liu, Y.; She, Y.; Pan, N.; Miao, Y.; Wang, X.; Liu, X.; Zhang, J.; Gao, W.; et al. Interlayer Donor-Acceptor Pair Excitons in MoSe2/WSe2Moiré Heterobilayer. *Nat. Commun.* **2023**, *14*, 5766.

(12) Tan, Q.; Lai, J.; Liu, X.; Guo, D.; Xue, Y.; Dou, X.; Sun, B.; Deng, H.; Tan, P.; Aharonovich, I.; Gao, W.; Zhang, J. Donor-

Acceptor Pair Quantum Emitters in Hexagonal Boron Nitride. *Nano Lett.* **2022**, *22*, 1331−1337.

(13) Bilgin, A.; Hammock, I. N.; Estes, J.; Jin, Y.; Bernien, H.; High, A. A.; Galli, G. Donor-Acceptor Pairs in Wide-Bandgap Semiconductors for Quantum Technology Applications. *npj. Comput. Mater.* **2024**, *10*, 7.

(14) Bradley, C. E.; Randall, J.; Abobeih, M. H.; Berrevoets, R. C.; Degen, M. J.; Bakker, M. A.; Markham, M.; Twitchen, D. J.; Taminiau, T. H. A Ten-Qubit Solid-State Spin Register with Quantum Memory up to One Minute. *Phys. Rev. X* **2019**, *9*, 031045.

(15) Neumann, P.; Kolesov, R.; Naydenov, B.; Beck, J.; Rempp, F.; Steiner, M.; Jacques, V.; Balasubramanian, G.; Markham, M. L.; Twitchen, D. J.; Pezzagna, S.; Meijer, J.; Twamley, J.; Jelezko, F.; Wrachtrup, J. Quantum Register Based on Coupled Electron Spins in a Room-Temperature Solid. *Nat. Phys.* **2010**, *6*, 249−253.

(16) de la Fuente, O R.; González-Barrio, M A; Navarro, V; Pabon, B M; Palacio, I; Mascaraque, A Surface Defects and Their Influence on Surface Properties. *J. Phys.:Condens. Matter* **2013**, *25*, 484008.

(17) Zhu, Y.; Yu, V. W.; Galli, G. First-Principles Investigation of Near-Surface Divacancies in Silicon Carbide. *Nano Lett.* **2023**, *23*, 11453−11460.

(18) Kaviani, M.; Deák, P.; Aradi, B.; Frauenheim, T.; Chou, J.; Gali, A. Proper Surface Termination for Luminescent Near-Surface NV Centers in Diamond. *Nano Lett.* **2014**, *14*, 4772−4777.

(19) Pasternak, D. G.; Romshin, A. M.; Bagramov, R. H.; Galimov, A. I.; Toropov, A. A.; Kalashnikov, D. A.; Leong, V.; Satanin, A. M.; Kudryavtsev, O. S.; Gritsienko, A. V.; Chernev, A. L.; Filonenko, V. P.; Vlasov, I. I. Donor−Acceptor Recombination Emission in Hydrogen-Terminated Nanodiamond. *Adv. Quantum Technol.* **2025**, *8*, 2400263.

(20) Devaty, R. P.; Choyke, W. J. Optical Characterization of Silicon Carbide Polytypes. *Phys. Status Solidi A* **1997**, *162*, 5−38.

(21) Levinshtein, M. E.; Rumyantsev, S. L.; Shur, M. S. *Properties of Advanced Semiconductor Materials: GaN, AIN, InN, BN, SiC, SiGe*; John Wiley & Sons Inc., New York, 2001.

(22) Trabada, D. G.; Flores, F.; Ortega, J. Hydrogenation of Semiconductor Surfaces: Si-Terminated Cubic SiC(100) Surfaces. *Phys. Rev. B* **2009**, *80*, 075307.

(23) Shinohara, M.; Katagiri, T.; Iwatsuji, K.; Matsuda, Y.; Fujiyama, H.; Kimura, Y.; Niwano, M. Oxidation of the Hydrogen Terminated Silicon Surfaces by Oxygen Plasma Investigated by In-Situ Infrared Spectroscopy. *Thin Solid Films* **2005**, *475*, 128−132.

(24) Yan, D.; Huang, H.; Huang, Y.; Yang, H.; Duan, N. Study on OH Radical Oxidation of 4H-SiC in Plasma Based on ReaxFF Molecular Dynamics Simulation. *J. Mol. Liq.* **2024**, *400*, 124573.

(25) Rashid, M.; Tiwari, A. K.; Goss, J. P.; Rayson, M. J.; Briddon, P. R.; Horsfall, A. B. Surface-State Dependent Optical Properties of OH-, F-, and H-terminated 4H-SiC Quantum Dots. *Phys. Chem. Chem. Phys.* **2016**, *18*, 21676−21685.

(26) Sayou Ngomsi, C. A.; Joshi, T.; Dev, P. Optimum surface-passivation schemes for near-surface spin defects in silicon carbide. *Phys. Rev. Mater.* **2024**, *8*, 056202.

(27) Polking, M. J.; Dibos, A. M.; de Leon, N. P.; Park, H. Improving Defect-Based Quantum Emitters in Silicon Carbide via Inorganic Passivation. *Adv. Mater.* **2018**, *30*, 1704543.

(28) Kuwabara, H.; Yamada, S.; Tsunekawa, S. Radiative Recombination in $\beta$-SiC Doped with Boron. *J. Lumin.* **1976**, *12−13*, 531−536.

(29) Freitas, J. A., Jr.; Bishop, S. G.; Nordquist, P. E. R.; Gipe, M. L. Donor Binding Energies Determined From Temperature Dependence of Photoluminescence Spectra in Undoped and Aluminum-Doped Beta SiC Films. *Appl. Phys. Lett.* **1988**, *52*, 1695−1697.

(30) Freitas, J. A., Jr.; Klein, P. B.; Bishop, S. G. Optical Studies of Donors and Acceptors in Cubic SiC. *Mater. Sci. Eng., B* **1992**, *11*, 21−25.

(31) Rao, M. V.; Griffiths, P.; Holland, O. W.; Kelner, G.; Freitas, J. A., Jr.; Simons, D. S.; Chi, P. H.; Ghezzo, M. Al and B Ion-Implantations in 6H and 3C-SiC. *J. Appl. Phys.* **1995**, *77*, 2479−2485.

(32) Perdew, J.; Burke, K.; Ernzerhof, M. Generalized Gradient Approximation Made Simple. *Phys. Rev. Lett.* **1996**, *77*, 3865−3868.

(33) Heyd, J.; Scuseria, G.; Ernzerhof, M. Hybrid Functionals Based on a Screened Coulomb Potential. *J. Chem. Phys.* **2003**, *118*, 8207−8215.

(34) Zheng, H.; Govoni, M.; Galli, G. Dielectric-Dependent Hybrid Functionals for Heterogeneous Materials. *Phys. Rev. Mater.* **2019**, *3*, 073803.

(35) Cipriano, L. A.; Di Liberto, G.; Tosoni, S.; Pacchioni, G. Quantum Confinement in Group III−V Semiconductor 2D Nanostructures. *Nanoscale* **2020**, *12*, 17494−17501.

(36) Liu, D. Density Functional Analysis of Key Energetics in Metal Homoepitaxy: Quantum Size Effects in Periodic Slab Calculations. *Phys. Rev. B* **2010**, *81*, 035415.

(37) Cannuccia, E.; Gali, A. Thermal Evolution of Silicon Carbide Electronic Bands. *Phys. Rev. Mater.* **2020**, *4*, 014601.

(38) Monserrat, B.; Needs, R. J. Comparing Electron-Phonon Coupling Strength in Diamond, Silicon, and Silicon Carbide: First-Principles Study. *Phys. Rev. B* **2014**, *89*, 214304.

(39) Ruhoff, P. T. Recursion Relations for Multi-Dimensional Franck-Condon Overlap Integrals. *Chem. Phys.* **1994**, *186*, 355−374.

(40) Alkauskas, A.; Lyons, J. L.; Steiauf, D.; Van de Walle, C. G. First-Principles Calculations of Luminescence Spectrum Line Shapes for Defects in Semiconductors: The Example of GaN and ZnO. *Phys. Rev. Lett.* **2012**, *109*, 267401.

(41) Alkauskas, A.; McCluskey, M. D.; Van de Walle, C. G. Tutorial: Defects in Semiconductors - Combining Experiment and Theory. *J. Appl. Phys.* **2016**, *119*, 181101.

(42) Jin, Y.; Govoni, M.; Wolfowicz, G.; Sullivan, S. E.; Heremans, F. J.; Awschalom, D. D.; Galli, G. Photoluminescence Spectra of Point Defects in Semiconductors: Validation of First-Principles Calculations. *Phys. Rev. Mater.* **2021**, *5*, 084603.

(43) Gali, A. Recent Advances in the Ab Initio Theory of Solid-State Defect Qubits. *Nanophotonics* **2023**, *12*, 359−397.

(44) Thomas, D. G.; Gershenzon, M.; Trumbore, F. A. Pair Spectra and 'Edge' Emission in Gallium Phosphide. *Phys. Rev.* **1964**, *133*, A269−A279.

(45) Dean, P. J. Bound Excitons and Donor−Acceptor Pairs in Natural and Synthethic Diamond. *Phys. Rev.* **1965**, *139*, A588−A602.

(46) Ziemelis, U. O.; Parsons, R. R. Sharp Line Donor-Acceptor Pair Luminescence in Silicon. *Can. J. Phys.* **1981**, *59*, 784−801.

(47) Williams, F. Donor—Acceptor Pairs in Semiconductors. *Phys. Status Solidi B* **1968**, *25*, 493−512.

(48) Schmid, W.; Nieper, U.; Weber, J. Donor-Acceptor Pair Spectra in Si:In LPE-Layers. *Solid State Commun.* **1983**, *45*, 1007−1011.

(49) Huang, K.; Rhys, A. Theory of Light Absorption and Non-Radiative Transitions in F-centers. *Proc. R. Soc. London, Ser. A* **1950**, *204*, 406−423.

(50) Alkauskas, A.; Buckley, B. B.; Awschalom, D. D.; Van de Walle, C. G. First-Principles Theory of the Luminescence Lineshape for the Triplet Tansition in Diamond NV Centres. *New J. Phys.* **2014**, *16*, 073026.

(51) Joshi, T.; Dev, P. Site-Dependent Properties of Quantum Emitters in Nanostructured Silicon Carbide. *PRX Quantum* **2022**, *3*, 020325.

(52) Spindlberger, L.; Csóré, A.; Thiering, G.; Putz, S.; Karhu, R.; Ul Hassan, J.; Son, N. T.; Fromherz, T.; Gali, A.; Trupke, M. Optical Properties of Vanadium in 4H Silicon Carbide for Quantum Technology. *Phys. Rev. Appl.* **2019**, *12*, 014015.

(53) Dombrowski, K. F.; Kaufmann, U.; Kunzer, M.; Maier, K.; Schneider, J.; Shields, V. B.; Spencer, M. G. Deep Donor State of Vanadium in Cubic Silicon Carbide (3C-SiC). *Appl. Phys. Lett.* **1994**, *65*, 1811−1813.

(54) Kunzer, M.; Müller, H. D.; Kaufmann, U. Magnetic Circular Dichroism and Site-Selective Optically Detected Magnetic Resonance of the Deep Amphoteric Vanadium Impurity in 6H-SiC. *Phys. Rev. B* **1993**, *48*, 10846−10854.

(55) Kaufmann, B.; Dörnen, A.; Ham, F. S. Crystal-Field Model of Vanadium in 6H Silicon Carbide. *Phys. Rev. B* **1997**, *55*, 13009–13019.
(56) Baur, J.; Kunzer, M.; Schneider, J. Transition Metals in SiC Polytypes, As Studied by Magnetic Resonance Techniques. *Phys. Status Solidi A* **1997**, *162*, 153–172.
(57) Ivády, V.; Abrikosov, I. A.; Janzén, E.; Gali, A. Role of Screening in the Density Functional Applied to Transition-Metal Defects in Semiconductors. *Phys. Rev. B* **2013**, *87*, 205201.
(58) Mitchel, W. C.; Mitchell, W. D.; Landis, G.; Smith, H. E.; Lee, W.; Zvanut, M. E. Vanadium Donor and Acceptor Levels in Semi-Insulating 4H- and 6H-SiC. *J. Appl. Phys.* **2007**, *101*, 013707.
(59) Spaldin, N. A. A Beginners Guide to the Modern Theory of Polarization. *J. Solid State Chem.* **2012**, *195*, 2–10.
(60) Pizzi, G.; Vitale, V.; Arita, R.; Blügel, S.; Freimuth, F.; Géranton, G.; Gibertini, M.; Gresch, D.; Johnson, C.; Koretsune, T.; Ibañez-Azpiroz, J.; Lee, H.; Lihm, J.; Marchand, D.; Marrazzo, A.; Mokrousov, Y.; Mustafa, J. I.; Nohara, Y.; Nomura, Y.; Paulatto, L.; et al. Wannier90 As a Community Code: New Features and Applications. *J. Phys.:Condens. Matter* **2020**, *32*, 165902.
(61) Govoni, M.; Galli, G. Large Scale GW Calculations. *J. Chem. Theory Comput.* **2015**, *11*, 2680–2696.
(62) Yu, V. W.; Govoni, M. GPU Acceleration of Large-Scale Full-Frequency GW Calculations. *J. Chem. Theory Comput.* **2022**, *18*, 4690–4707.
(63) Shields, B. J.; Unterreithmeier, Q. P.; de Leon, N. P.; Park, H.; Lukin, M. D. Efficient Readout of a Single Spin State in Diamond via Spin-to-Charge Conversion. *Phys. Rev. Lett.* **2015**, *114*, 136402.
(64) Giannozzi, P.; Baseggio, O.; Bonfà, P.; Brunato, D.; Car, R.; Carnimeo, I.; Cavazzoni, C.; de Gironcoli, S.; Delugas, P.; Ruffino, F. F.; Ferretti, A.; Marzari, N.; Timrov, I.; Urru, A.; Baroni, S. Quantum ESPRESSO Toward the Exascale. *J. Chem. Phys.* **2020**, *152*, 154105.
(65) Giannozzi, P.; Andreussi, O.; Brumme, T.; Bunau, O.; Nardelli, M. B.; Calandra, M.; Car, R.; Cavazzoni, C.; Ceresoli, D.; Cococcioni, M.; Colonna, N.; Carnimeo, I.; Dal Corso, A.; de Gironcoli, S.; Delugas, P.; DiStasio, R. A., Jr.; Ferretti, A.; Floris, A.; Fratesi, G.; Fugallo, G.; et al. Advanced Capabilities for Materials Modelling with Quantum ESPRESSO. *J. Phys.:Condens. Matter* **2017**, *29*, 465901.
(66) Giannozzi, P.; Baroni, S.; Bonini, N.; Calandra, M.; Car, R.; Cavazzoni, C.; Ceresoli, D.; Chiarotti, G. L.; Cococcioni, M.; Dabo, I.; Dal Corso, A.; de Gironcoli, S.; Fabris, S.; Fratesi, G.; Gebauer, R.; Gerstmann, U.; Gougoussis, C.; Kokalj, A.; Lazzeri, M.; Martin-Samos, L.; et al. Quantum ESPRESSO: A Modular and Open-Source Software Project for Quantum Simulations of Materials. *J. Phys.:Condens. Matter* **2009**, *21*, 395502.
(67) Schlipf, M.; Gygi, F. Optimization Algorithm for the Generation of ONCV Pseudopotentials. *Comput. Phys. Commun.* **2015**, *196*, 36–44.
(68) Hamann, D. R. Optimized Norm-Conserving Vanderbilt Pseudopotentials. *Phys. Rev. B* **2013**, *88*, 085117.
(69) Bengtsson, L. Dipole correction for surface supercell calculations. *Phys. Rev. B* **1999**, *59*, 12301.